\documentclass[conference]{IEEEtran}
\IEEEoverridecommandlockouts

\usepackage{cite}
\usepackage{amsmath,amssymb,amsfonts}
\usepackage{tabularx}
\usepackage{multirow}
\usepackage[dvipdfmx]{graphicx}
\usepackage{textcomp}
\usepackage[dvipdfmx]{color}
\usepackage{here}
\usepackage{url}
\usepackage{breakurl}
\usepackage{listings}
\usepackage[dvipsnames]{xcolor}
\usepackage{listings} 
\lstdefinestyle{customcpp}{
  language=C++,
  basicstyle=\ttfamily\scriptsize,
  keywordstyle=\color{blue}\bfseries,
  commentstyle=\color{gray},
  stringstyle=\color{red},
  numberstyle=\tiny\color{gray},
  numbers=left, 
  stepnumber=1,
  breaklines=true, 
  frame=single, 
  xleftmargin=2em,
  tabsize=2
}
\def\BibTeX{{\rm B\kern-.05em{\sc i\kern-.025em b}\kern-.08em
    T\kern-.1667em\lower.7ex\hbox{E}\kern-.125emX}}
\begin{document}

\title{Exploring the Versal AI Engine for \\ 3D Gaussian Splatting
}

\author{\IEEEauthorblockN{Kotaro Shimamura, Ayumi Ohno, Shinya Takamaeda-Yamazaki}
\IEEEauthorblockA{\textit{The University of Tokyo} \\
\{kotaro, ayumi0130ohno, shinya\} @is.s.u-tokyo.ac.jp}
}

\maketitle

\begin{abstract}
  Dataflow-oriented spatial architectures are the emerging paradigm for higher computation performance and efficiency.
  AMD Versal AI Engine is a commercial spatial architecture consisting of tiles of VLIW processors supporting SIMD operations arranged in a two-dimensional mesh.
  The architecture requires the explicit design of task assignments and dataflow configurations for each tile to maximize performance, demanding advanced techniques and meticulous design.
  However, a few works revealed the performance characteristics of the Versal AI Engine through practical workloads.
  In this work, we provide the comprehensive performance evaluation of the Versal AI Engine using Gaussian feature computation in 3D Gaussian splatting as a practical workload, and we then propose a novel dedicated algorithm to fully exploit the hardware architecture.
  The computations of 3D Gaussian splatting include matrix multiplications and color computations utilizing high-dimensional spherical harmonic coefficients.
  These tasks are processed efficiently by leveraging the SIMD capabilities and their instruction-level parallelism.
  Additionally, pipelined processing is achieved by assigning different tasks to individual cores, thereby fully exploiting the spatial parallelism of AI Engines.
  The proposed method demonstrated a 226-fold throughput increase in simulation-based evaluation, outperforming a naive approach.
  These findings provide valuable insights for application development that effectively harnesses the spatial and architectural advantages of AI Engines.
\end{abstract}


\section{Introduction}

Historically, improvements in performance were driven by transistor scaling, as predicted by Moore's Law; however, physical limitations and escalating manufacturing costs have recently hindered these conventional approaches \cite{theis2017end}. 
As a result, hardware design has shifted toward exploiting parallelism and diversifying system architectures.

Against this backdrop, AMD (formerly Xilinx) introduced the Versal Adaptive Compute Acceleration Platform (Versal ACAP) \cite{gaide2019xilinx}.
This platform advances traditional FPGA technology by integrating a processing subsystem (PS), Programmable Logic (PL), and the AI Engines into a single device, forming a heterogeneous architecture.
Among its key features, AI Engines stand out as a dedicated unit for high-performance computation.

The AI Engines consist of an array of tiles arranged in a two-dimensional mesh, with each tile containing a 7-way Very Long Instruction Word (VLIW) processor, local memory, and interconnects.
Additionally, by leveraging multiple levels of parallelism—such as SIMD, instruction-level parallelism (ILP), task-level parallelism, and data-level parallelism—the AI Engines achieve both high performance and flexibility.

This study aims to comprehensively evaluate the performance of AI Engines using Gaussian feature computation in 3D Gaussian Splatting (3DGS) as the target workload. 
In conjunction with this evaluation, we propose a novel hardware algorithm that fully leverages the architecture of AI Engines.

3DGS is a technique for constructing three-dimensional scenes from large sets of camera images. 
Unlike neural networks, it represents shapes and appearances using Gaussian-based point clouds, making it well-suited for real-time processing \cite{kerbl20233d}. 
This method requires processing tens of thousands to millions of Gaussians, and leveraging the high data parallelism of AI Engines can significantly enhance performance. 
Additionally, since 3DGS involves diverse computational tasks such as matrix operations and spherical harmonic coefficient calculations, further optimization can be achieved by exploiting task-level parallelism and SIMD capabilities.
For these reasons, we determined that 3DGS is a suitable application for evaluating the parallel processing capabilities of AI Engines and selected it as the focus of this study.

To maximize the performance of AI Engines, this study focuses on two key aspects: In-tile optimization and spatial parallelism. 
First, for In-tile optimization, we implement vector optimizations to leverage VLIW and SIMD capabilities while also performing task partitioning to enable efficient execution of task pipelines. 
Second, in terms of spatial parallelism, we explore efficient mapping strategies within AI Engine Array to construct an architecture that maximizes computational resource utilization.
Finally, we validate the effectiveness of the hardware implementation through detailed performance analysis using the AI Engine multi-threaded simulator and system-level evaluation on the VCK190 platform.

The primary contribution of this study is the development of a highly efficient hardware algorithm that fully exploits the parallel processing characteristics and tile-based architecture of AI Engines. 
Experimental results demonstrate that the proposed method achieves up to a 226-fold increase in throughput compared to a naive implementation. 
Furthermore, the insights gained from this study are not limited to 3DGS but can also be applied to other highly parallel computing applications such as AI inference and image processing, providing valuable guidelines for the future design of tile-based architectures.

\section{Background}

\subsection{Versal ACAP}

Versal ACAP is a heterogeneous architecture that integrates a Processing Subsystem (PS), Programmable Logic (PL), and AI Engines into a single device.

The VCK190 platform used in this study is equipped with AI Engine Array, which features a \textbf{\(50 \times 8\)} two-dimensional mesh structure.  
Each tile in the array consists of a 7-way Very Long Instruction Word (VLIW) vector processor, local memory (32 KB), and instruction memory (16 KB).  
By assigning independent tasks to each tile, the architecture enables high parallelism, thereby enhancing computational throughput.  
The 7-way VLIW vector processor can issue up to seven instructions per clock cycle and execute multiply-accumulate (MAC) operations on up to 128 elements simultaneously, achieving a peak performance of 256 operations per clock cycle.  
AI Engine programming is supported in high-level languages such as C/C++, allowing developers to leverage AI Engine-specific APIs and intrinsics to maintain code readability while optimizing performance \cite{xilinx_aie_api, xilinx_aie_intrinsics}.  

The AI Engines provide multiple communication methods for efficient data exchange between tiles. 
In this study, two communication methods are utilized to optimize performance.
The first method is the \textbf{Window Interface}, which allows direct access to the local memory of adjacent tiles. Each tile is equipped with two 256-bit load units and one 256-bit store unit, enabling up to two load operations and one store operation per clock cycle.
The second method is the \textbf{Stream Interface}, which facilitates data transmission between distant tiles or between the AI Engines and the PL using AXI4-Stream. 
Each tile has two input ports and two output ports, supporting data transfer at 32 bits per cycle or 128 bits every four cycles through FIFO-based communication. 
This approach is advantageous in applications requiring flexible data flow configurations or when direct tile-to-tile connections are not feasible.

In this study, \textbf{PL Interface Tile (PLIO)} is used for data communication between the AI Engines and external units such as the PL, PS, and NoC.  
PLIO facilitates high-speed data transfer between the AI Engine and the PL via the AXI4-Stream interface.  
On the VCK190 platform, the system achieves a total bandwidth of 1.0 TB/s for AI Engine to PL transfers and 1.3 TB/s for PL to AI Engines transfers, contributing to overall system performance enhancement \cite{amd_datasheet}.

\subsection{3D Gaussian Splatting (3DGS)}
\label{sec:3dgs}

3DGS is a cutting-edge method for 3D reconstruction, similar to Neural Radiance Fields (NeRF)~\cite{mildenhall2021nerf}, but instead of relying on voxel grids or neural networks, it represents scene geometry and appearance using numerous Gaussian distributions~\cite{kerbl20233d}.  

Each Gaussian is defined by its position \(\mathbf{p}_w \in \mathbb{R}^3\), rotation \(\mathbf{q} \in \mathbb{R}^4\), scale \(\mathbf{s} = (s_x, s_y, s_z) \in \mathbb{R}^3\), spherical harmonic coefficients \(\mathbf{sh} \in \mathbb{R}^{48}\) for view-dependent color, and opacity \(\alpha \in \mathbb{R}\).  

During rendering, each Gaussian is projected onto the 2D image plane based on camera parameters, yielding its projection coordinates \(\mathbf{u} \in \mathbb{R}^2\), 2D covariance matrix $\mathbf{\Sigma'} \in \mathbb{R}^{2\times2}$, and view-dependent color \(\mathbf{c} \in \mathbb{R}^3\).  

The projection coordinates \(\mathbf{u}\) are obtained by transforming \(\mathbf{p}_w\) using the camera rotation matrix \(\mathbf{R}_{cw}\) and intrinsic parameters.  
The spatial uncertainty of each Gaussian is represented by its 3D covariance matrix \(\mathbf{\Sigma}\), which is computed using the rotation matrix \(\mathbf{R}\) and the scale matrix \(\mathbf{S} = \text{diag}(s_x^2, s_y^2, s_z^2)\), both derived from the Gaussian parameters \(\mathbf{q}\) and \(\mathbf{s}\):  
\begin{equation}
\mathbf{\Sigma} = \mathbf{R} \mathbf{S} \mathbf{R}^{T}.
\label{eq:covariance_matrix_3d}
\end{equation}
To project this uncertainty onto the 2D image plane, the 2D covariance matrix is derived as follows:
\begin{equation}
\mathbf{\Sigma'} = \mathbf{J} \mathbf{R}_{cw} \mathbf{\Sigma} \mathbf{R}_{cw}^{T} \mathbf{J}^{T}.
\label{eq:covariance_matrix_2d}
\end{equation}
where $\mathbf{J}$ represents the Jacobian matrix.
The view-dependent color \(\mathbf{c}(\mathbf{p}_w, \mathbf{r})\) is computed using spherical harmonic coefficients \(\mathbf{sh}\) and the viewing direction \(\mathbf{r}\):  
\begin{equation}
\mathbf{c}(\mathbf{p}_w, \mathbf{r}) = \sum_{l=0}^{3} \sum_{m=-l}^{l} \mathbf{sh}_{lm} Y_{lm}(\mathbf{r}),
\label{eq:sh_color}
\end{equation}
where \(Y_{lm}(\mathbf{r})\) are the spherical harmonic basis functions, and \(\mathbf{sh}_{lm}\) are the corresponding coefficients for each Gaussian.

Once these attributes are computed, 3DGS employs a volume rendering approach that sorts Gaussians based on viewpoint, blends their contributions along each pixel's line of sight, and accumulates opacity to account for occlusion.  
This enables efficient and visually coherent rendering without the need for large-scale neural computations.

In the training phase, 3DGS estimates camera parameters and a 3D point cloud via Structure-from-Motion (SfM)~\cite{schonberger2016structure}, converts each point into an initial Gaussian, and refines its parameters through differentiable Gaussian splatting.  
The optimization process is guided by a loss function combining pixel-wise and perceptual metrics, such as an $L_1$ term and a D-SSIM term, ensuring the synthesized images match ground-truth views.  
Additional procedures dynamically adjust the number of Gaussians by replicating those in high-detail regions and removing low-opacity ones, balancing density and accuracy.  
The final optimized Gaussians enable real-time high-fidelity novel view synthesis with minimal computational overhead.

By avoiding large neural networks and instead utilizing a collection of Gaussian primitives, 3DGS efficiently processes large-scale 3D scenes, making it well suited for mobile and edge devices with limited computational resources.  
Moreover, its direct geometric representation allows intuitive modifications and extensions, establishing 3DGS as a compelling alternative to conventional mesh- or network-based methods for high-quality, efficient 3D reconstruction.

\section{System Architecture and Dataflow}

\begin{figure}[htbp]
  \centering
  \includegraphics[width=0.45\textwidth]{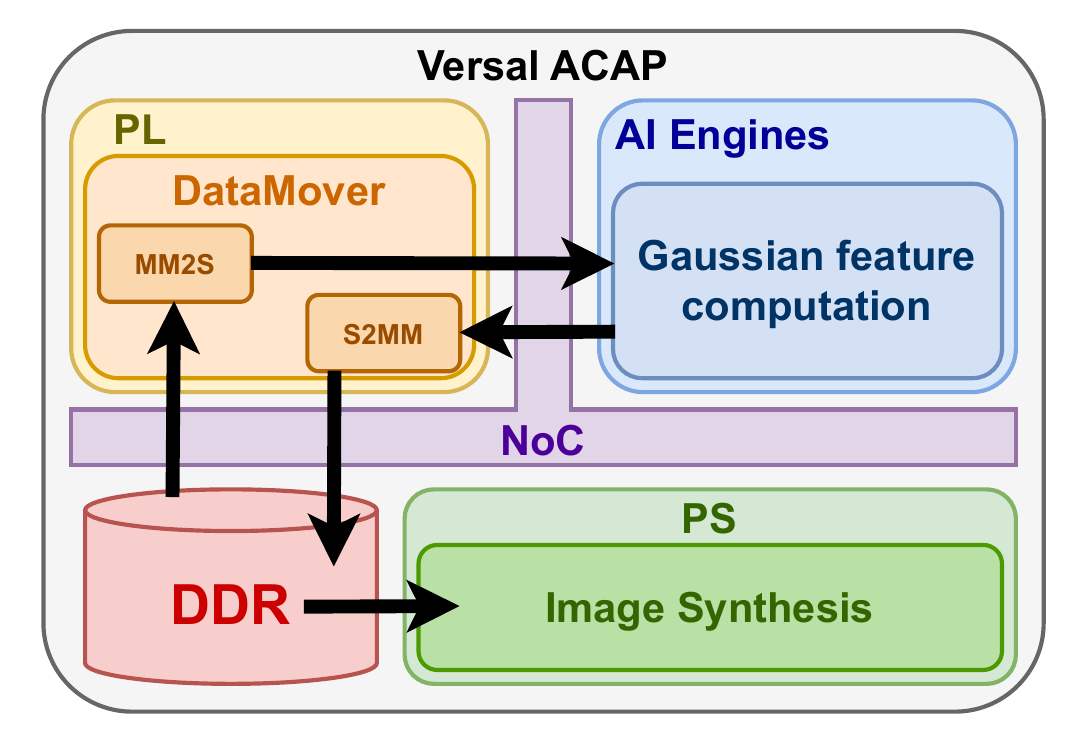} 
  \caption{System Architecture and Dataflow of 3DGS on Versal ACAP.}
  \label{fig:figure5}
\end{figure}

Figure~\ref{fig:figure5} illustrates the architecture of 3DGS implemented on the VCK190 platform.

Data transfer is managed by the DataMover module (MM2S and S2MM) implemented in the PL, facilitating Gaussian data exchange between DDR memory and AI Engines.  
AI Engines compute the features of each Gaussian received from the PL and transmit the necessary outputs for image generation back to the PL.  
By leveraging the parallel processing capabilities of AI Engines, large-scale Gaussian datasets can be processed efficiently with high throughput.

The computed features are written back to DDR memory, which the PS retrieves for image generation.
Additionally, the PS utilizes Xilinx Runtime (XRT) to manage the control of the PL and AI Engines, oversee kernel execution, perform memory mapping, and optimize data transfers.

This implementation fully utilizes the heterogeneous computing capabilities of AI Engines.

\section{Hardware Algorithm for Feature Computation using AI Engines}

\begin{figure}[htbp]
  \centering
  \includegraphics[width=0.45\textwidth]{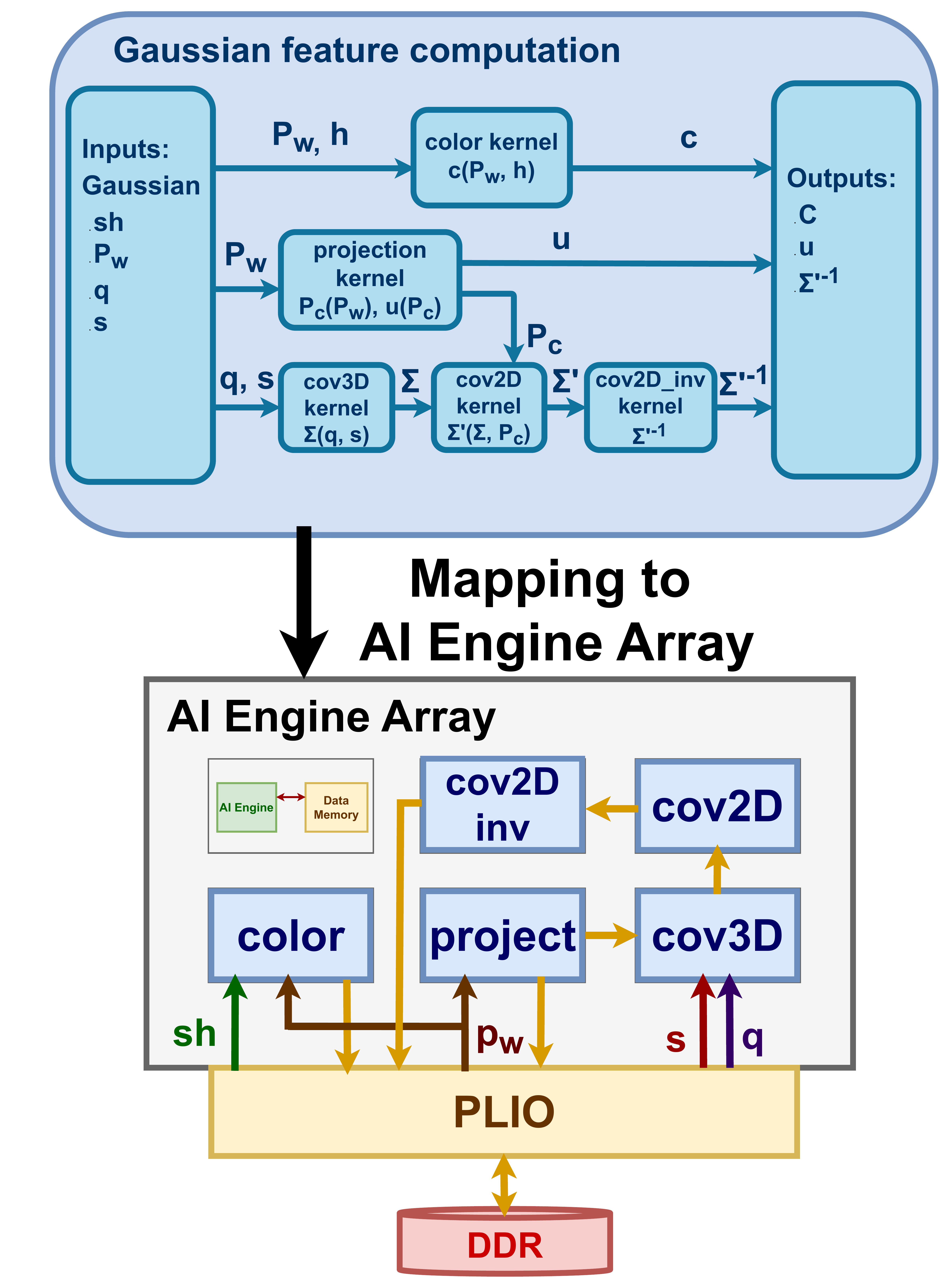} 
  \caption{Overall Gaussian feature computation and its naive mapping to AI Engine Array.}
  \label{fig:figure6}
\end{figure}

This section explores multiple hardware algorithms to implement feature computation on AI Engines.

\subsection{Naive Approach}

First, we examine a naive approach.  
Figure~\ref{fig:figure6} illustrates the processing flow of Gaussian feature computation and its naive mapping onto AI Engine Array.  
Each task in feature computation is implemented as an individual kernel and assigned to a single AI Engine tile.  
These tasks include small-scale matrix multiplications and color computations, but vector operations are not utilized within the kernels.  
Additionally, this implementation does not parallelize the feature computation units; instead, only a single unit is mapped onto AI Engine Array.  
This naive approach serves as a baseline before applying optimizations, providing a reference point for performance evaluation.

\subsection{In-tile Optimization Approach}

Compared to the naive implementation, an optimized approach is designed to maximize the performance of AI Engines. 
The following optimizations are applied:

\subsubsection{Vector Optimization}

To maximize the vector computation performance of AI Engines, specific optimizations were applied using dedicated APIs and intrinsic functions. 
This section describes the optimization approach with a focus on the computation of the 3D covariance matrix $\mathbf{\Sigma}$ as defined in Equation~\eqref{eq:covariance_matrix_3d}. 

In a naive approach, the computation is performed using a triple loop, as shown in Listing~\ref{alg:naive_cov3D}.

\begin{lstlisting}[style=customcpp, caption={Naive Computation of the 3D Covariance Matrix}, captionpos=b, label=alg:naive_cov3D]
for(int i = 0; i < 3; ++i) {
    for(int j = 0; j < 3; ++j) {
        temp[i * 3 + j] = 0;
        for(int k = 0; k < 3; ++k) {
            temp[i * 3 + j] += R[i * 3 + k] * S[k * 3 + j];
        }
    }
}

for(int i = 0; i < 3; ++i) {
    for(int j = 0; j < 3; ++j) {
        cov3D[i * 3 + j] = 0;
        for(int k = 0; k < 3; ++k) {
            cov3D[i * 3 + j] += temp[i * 3 + k] * R[j * 3 + k];
        }
    }
}
\end{lstlisting}

The optimized implementation leverages the AI Engine's vector APIs, enabling the use of the VLIW vector execution slots and supporting both SIMD and instruction-level parallelism (ILP). 

By applying the following transformations, vectorized operations can be efficiently achieved:

\[
\mathbf{\Sigma} = \mathbf{R} \cdot \text{diag}(s_x^2, s_y^2, s_z^2) \cdot \mathbf{R}^{T}
\]

\[
= 
\begin{bmatrix}
\mathbf{r}_1 \\
\mathbf{r}_2 \\
\mathbf{r}_3
\end{bmatrix}
\begin{bmatrix}
s_x^2 & 0 & 0 \\
0 & s_y^2 & 0 \\
0 & 0 & s_z^2
\end{bmatrix}
\begin{bmatrix}
\mathbf{r}_1^{T} & \mathbf{r}_2^{T} & \mathbf{r}_3^{T}
\end{bmatrix}
\]

\[
=
\begin{bmatrix}
\mathbf{r}_1 \cdot (\mathbf{r}_1 \odot \mathbf{s}) & \mathbf{r}_1 \cdot (\mathbf{r}_2 \odot \mathbf{s}) & \mathbf{r}_1 \cdot (\mathbf{r}_3 \odot \mathbf{s}) \\
\mathbf{r}_2 \cdot (\mathbf{r}_1 \odot \mathbf{s}) & \mathbf{r}_2 \cdot (\mathbf{r}_2 \odot \mathbf{s}) & \mathbf{r}_2 \cdot (\mathbf{r}_3 \odot \mathbf{s}) \\
\mathbf{r}_3 \cdot (\mathbf{r}_1 \odot \mathbf{s}) & \mathbf{r}_3 \cdot (\mathbf{r}_2 \odot \mathbf{s}) & \mathbf{r}_3 \cdot (\mathbf{r}_3 \odot \mathbf{s})
\end{bmatrix}.
\]

Since the 3D covariance matrix is symmetric, only the upper triangular elements need to be computed, reducing redundant calculations. 
The AI Engine's vectorized API is used for efficient element-wise multiplication, as shown in Listing~\ref{alg:vectorized_cov3D}:

\begin{lstlisting}[style=customcpp, caption={Vectorized Computation of the 3D Covariance Matrix}, captionpos=b, label=alg:vectorized_cov3D]
cov3D_1 = aie::mul(aie::mul(S, R1).to_vector(), R1);
cov3D_2 = aie::mul(aie::mul(S, R1).to_vector(), R2);
cov3D_3 = aie::mul(aie::mul(S, R1).to_vector(), R3);
cov3D_5 = aie::mul(aie::mul(S, R2).to_vector(), R2);
cov3D_6 = aie::mul(aie::mul(S, R2).to_vector(), R3);
cov3D_9 = aie::mul(aie::mul(S, R3).to_vector(), R3);
\end{lstlisting}

Thus, the optimized implementation minimizes computational overhead while maximizing the utilization of AI Engine's vector execution resources.

\subsubsection{Task Partitioning}

\begin{figure}[htbp]
  \centering
  \includegraphics[width=0.5\textwidth]{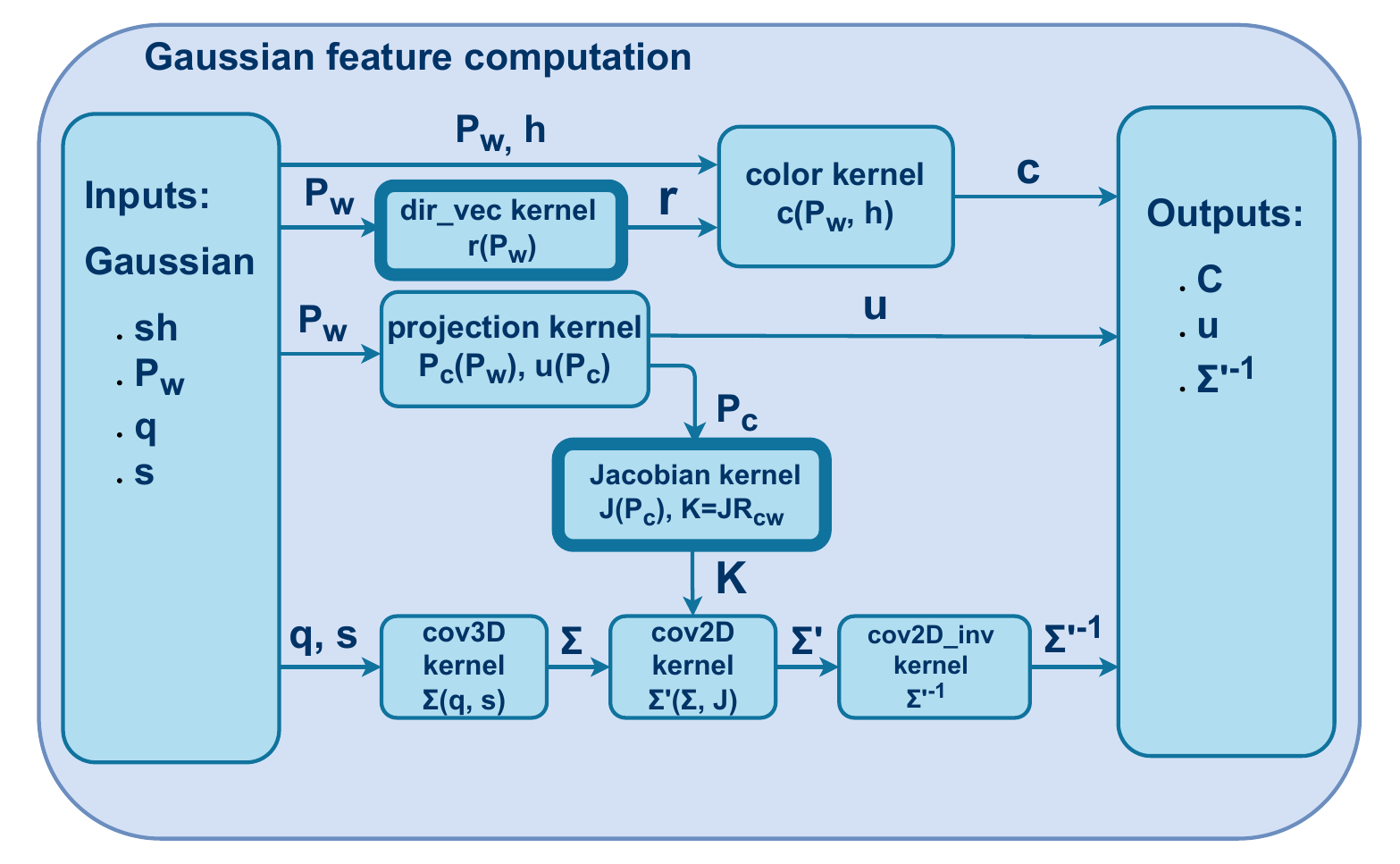} 
  \caption{Overview of Gaussian feature computation after task partitioning.}
  \label{fig:figure7}
\end{figure}

The primary objective of task partitioning is to efficiently structure the pipeline stages, ensuring smooth execution of the task pipeline. Additionally, this partitioning significantly impacts the utilization of spatial parallelism in AI Engine Array, as discussed later.

As shown in Figure~\ref{fig:figure7}, two computationally intensive kernels, color and cov2D kernels, are partitioned to optimize processing efficiency.

In the color kernel, the spherical harmonic coefficients $\mathbf{sh}$ have a size of 48, making the computation highly demanding. To mitigate this workload, a dedicated kernel called ray\_dir kernel is introduced to handle preprocessing, reducing the computational burden of the main color computation.

The computation of the 2D covariance matrix $\mathbf{\Sigma'}$ involves five matrix multiplications as defined in Equation~\eqref{eq:covariance_matrix_2d}. 

To reduce computational load, an intermediate variable $\mathbf{K}$ is introduced:
\begin{equation}
\mathbf{K} = \mathbf{J} \mathbf{R}_{cw}
\end{equation}

By precomputing and storing $\mathbf{K}$, redundant matrix multiplications in the computation of $\mathbf{\Sigma'}$ are eliminated. 
This partitioning strategy enhances computational efficiency and optimizes data reuse within AI Engine Array.

\subsection{Leveraging Spatial Parallelism in AI Engines}

\begin{figure}[t]
    \centering
    \includegraphics[width=0.45\textwidth]{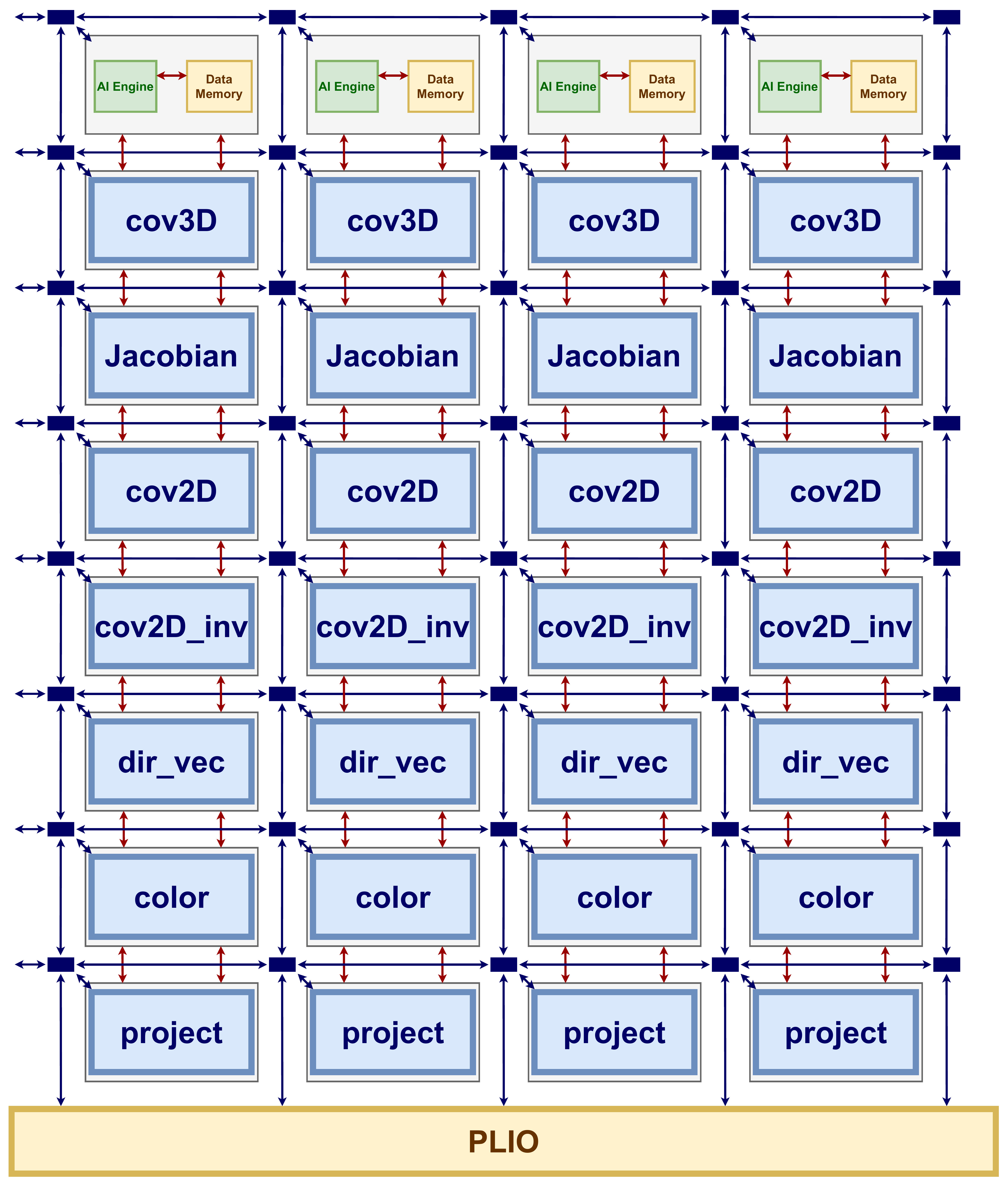} 
    \caption{Mapping of feature computation units in AI Engine tiles.}
    \label{fig:tile_mapping}
\end{figure}

To maximize the utilization of AI Engine tiles, this study adopts an approach that leverages spatial parallelism. 
The VCK190 platform consists of an array of 8 rows $\times$ 50 columns, totaling 400 AI Engine tiles. 
This method constructs multiple feature computation units and arranges them spatially to achieve parallel execution.
In kernel mapping, the number of PLIO streams is limited, necessitating appropriate task partitioning to maximize spatial parallelism. 
To address this, In-Tile optimization is applied to efficiently allocate kernels while considering PLIO constraints. Additionally, to minimize communication overhead, the distance from PLIO is kept as uniform as possible.
Considering these constraints, kernels are aligned in the column direction to optimize computation. Furthermore, task partitioning increases the number of kernels from 5 to 7, enabling the utilization of 7 out of 8 tiles per column for computation.
Figure~\ref{fig:tile_mapping} illustrates an example where feature computation units are configured in four parallel instances.
In this method, all 50 columns can be utilized, allowing a maximum of 50 parallel instances, resulting in the simultaneous use of 7 $\times$ 50 = 350 tiles.

\section{Experiment}

\subsection{Experimental Setup}
In this study, we evaluated the system design using the VCK190 platform and implemented the system with Vitis 2022.2.

The evaluation of the designed system was divided into two parts: evaluation of the AI Engines alone and evaluation of the entire system.

\subsubsection{Evaluation of the AI Engine}

The evaluation of the AI Engines alone was conducted using the AI Engine multi-threaded simulator. 
This simulator models the timing and resource usage of AI Engine array and simulates memory-related operations using transaction-level System C models for the NoC and DDR memory.  
Inputs and outputs for the simulator were provided by connecting input and output text files to PLIO. 
For evaluation, 100 Gaussian samples were randomly generated and used as input data. 
Using this simulator, cycle-accurate timing information for the application could be obtained.  
All AI Engine tiles were operated at 1.25 GHz.

\subsubsection{Evaluation of the Entire System}

The system-level evaluation was conducted through real-world measurements on the VCK190 platform.
For this study, we utilized the train set of the publicly available dataset \texttt{tandt\_db}, provided by the INRIA GraphDeco team.
This dataset is part of the 3D Gaussian Splatting project, and further details can be found in reference~\cite{inria_gaussian_splatting}.

Using the dataset and our custom training code, a set of Gaussians was generated.
A ground-truth image from the dataset was reconstructed using the trained Gaussian model and appropriately configured camera parameters.
During this process, the time required for image generation was measured.
The trained Gaussian model used for the evaluation contained 389,434 Gaussians.

Similar to the AI Engines evaluation, all AI Engine tiles were operated at 1.25 GHz during the system-level evaluation to ensure consistent performance conditions between the simulation and the real-world tests.

\subsection{Results and Discussion}

This section presents the simulation-based evaluation of the AI Engines alone and the system-wide performance evaluation on the VCK190 platform.

\subsubsection{Evaluation of the AI Engines}

\begin{figure}[htbp]
  \centering
  \includegraphics[width=0.46\textwidth]{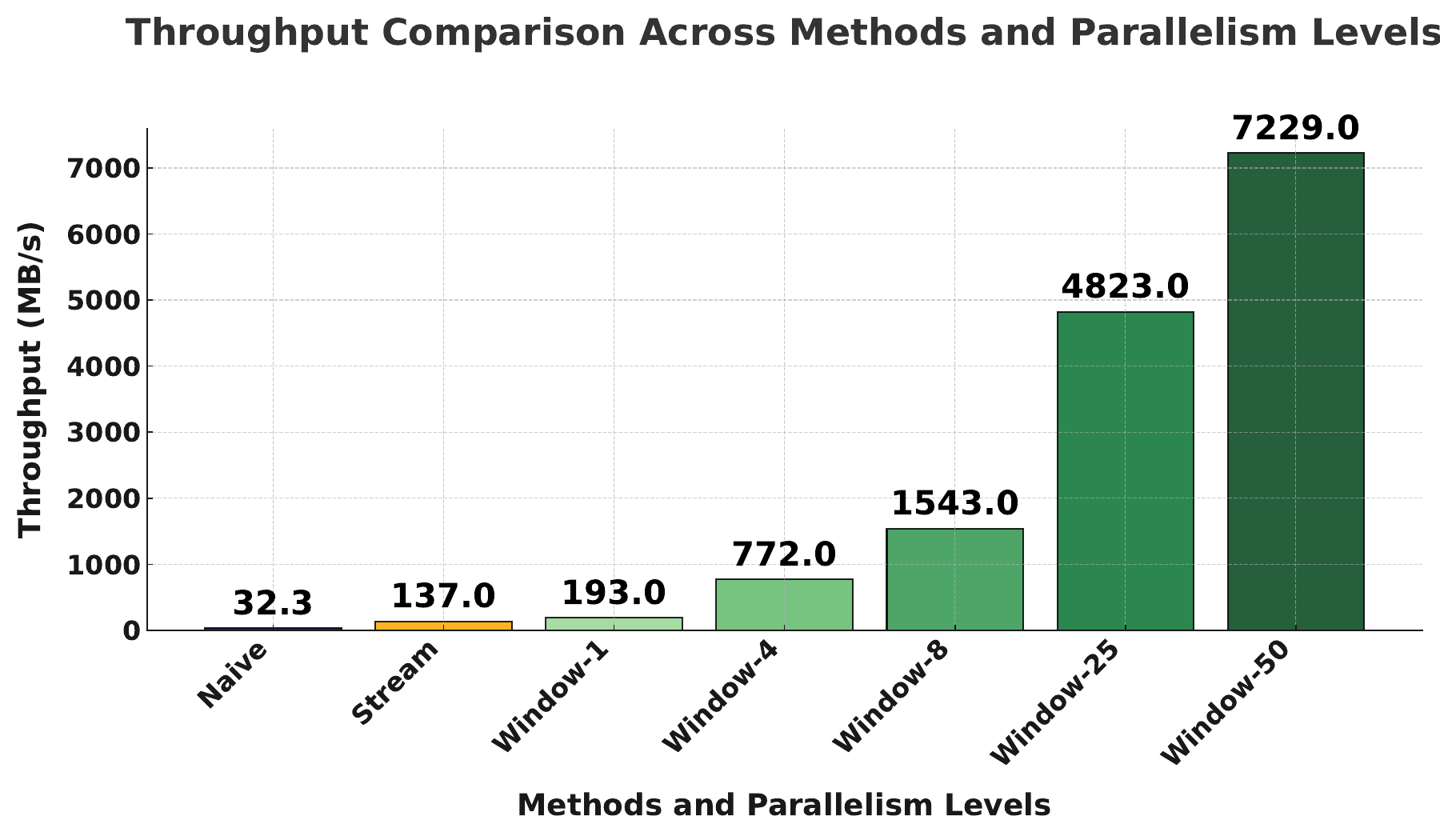}
  \caption{Throughput comparison across methods and parallelism levels.}
  \label{fig:throughput_comparison}
\end{figure}

\begin{table}[htbp]
  \centering
  \caption{Comparison of the number of cycles required by each kernel to compute the features of a single Gaussian.}
  \label{tab:kernel_performance_comparison}
  \scriptsize
  \renewcommand{\arraystretch}{1.2} 
  \setlength{\tabcolsep}{2pt} 
  \resizebox{\linewidth}{!}{ 
  \begin{tabular}{|c|c|c|c|c|c|c|c|c|}
    \hline
    \textbf{Methods} & \textbf{Metric} & \textbf{color} & \textbf{dir\_vec} & \textbf{cov2D} & \textbf{Jacobian} & \textbf{cov2D\_inv} & \textbf{projection} & \textbf{cov3D} \\
    \hline
    \multirow{2}{*}{Naive} 
      & Avg      & 1822 & -   & 1342 & -   & 1180 & 670 & 276 \\ 
      & min-max  & 1812-1861 & - & 1332-1381 & - & 1180-1181 & 670-671 & 276-277 \\ 
    \hline
    \multirow{2}{*}{Stream} 
      & Avg      & 433  & 262 & 225  & 135 & 230  & 79  & 210 \\ 
      & min-max  & 428-485  & 77-428  & 158-429  & 124-214  & 158-483  & 79-429  & 210-429 \\ 
    \hline
    \multirow{2}{*}{Window} 
      & Avg      & 371  & 83  & 184  & 130 & 57   & 89  & 194 \\ 
      & min-max  & 371-371  & 83-83  & 183-185  & 130-132  & 57-57  & 89-89  & 194-196 \\ 
    \hline
  \end{tabular}
  }
\end{table}

\begin{figure}[htbp]
  \centering
  \includegraphics[width=0.48\textwidth]{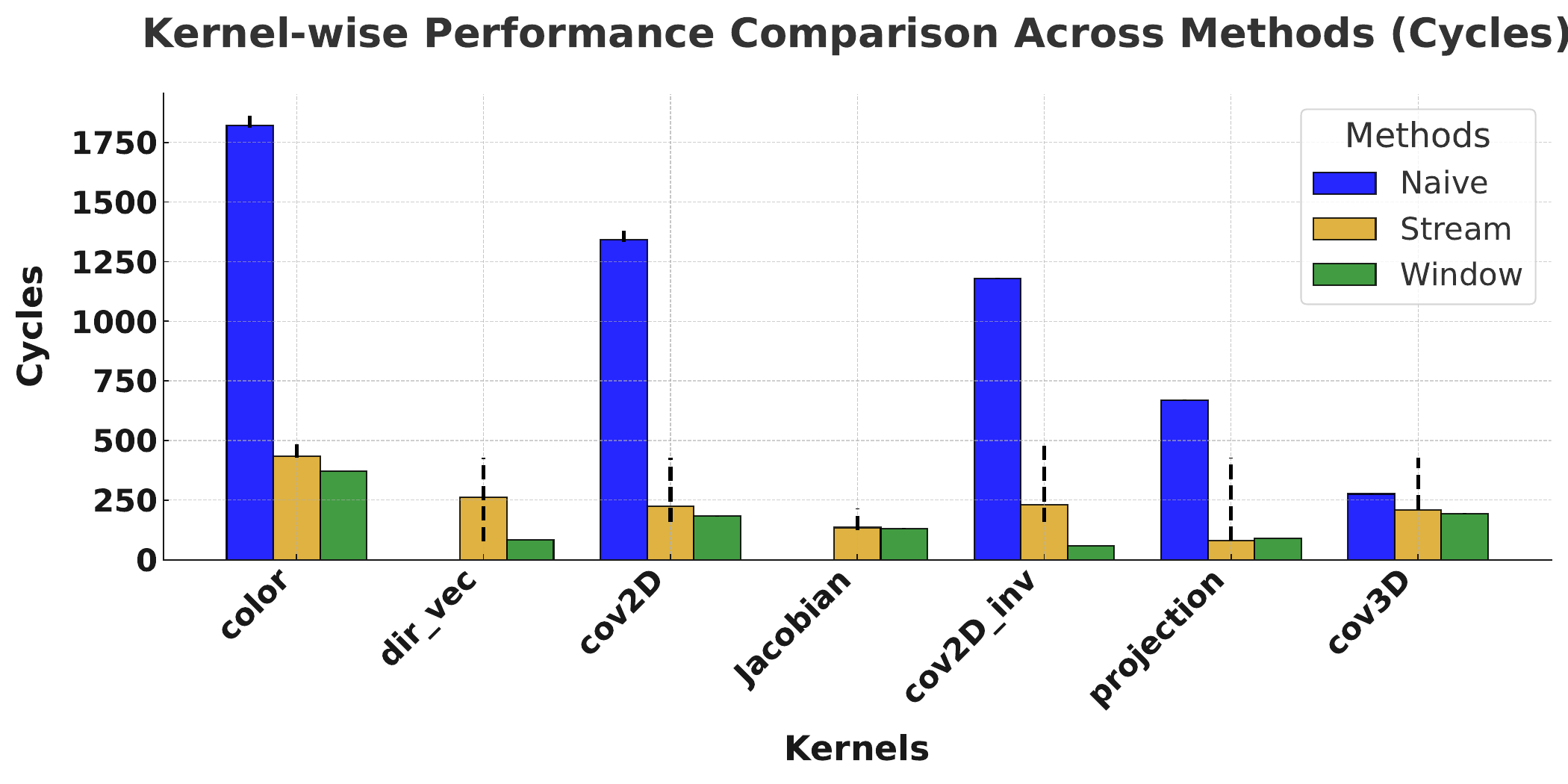}
  \caption{Comparison of the number of cycles required by each kernel to compute the features of a single Gaussian.}
  \label{fig:kernel_performance_comparison}
\end{figure}

As shown in Figure~\ref{fig:throughput_comparison}, the throughput (MB/s) for each method was measured using the AI Engine multi-threaded simulator.

The Naive method serves as a baseline implementation with no optimizations, while the other methods include In-tile optimizations. 
The Stream method utilizes the Stream interface for inter-AI Engines data transmission, whereas the Window method employs the Window interface. 
Furthermore, the Window method includes results measured with multiple feature computation units (parallelism degrees) of 1, 4, 8, 25, and 50.
Comparing Naive, Stream, and Window, it is evident that In-tile optimizations significantly improve throughput. 
Specifically, the Window method achieves approximately 6x higher throughput compared to the Naive method.

The superior performance of the Window method over the Stream method can be attributed to the difference in interface bandwidth. 
The Stream interface transmits only 32 bits of data per cycle, whereas the Window interface supports up to 256 bits per cycle. 
This difference reduces data transfer overhead, thereby enhancing the performance of the Window method.

Moreover, increasing the parallelism degree in the Window method improves throughput. 
Between parallelism degrees of 1 and 25, performance increases nearly linearly. 
However, beyond 25, the performance gain deviates from linear scaling, resulting in lower-than-expected improvements. 
This degradation is likely due to increased data transfer demands within the AI Engines, leading to frequent data stalls that hinder performance.

On the other hand, Table~\ref{tab:kernel_performance_comparison} presents the performance of individual kernels, showing the average, minimum, and maximum cycle counts required to compute the features of a single Gaussian. Lower cycle counts indicate faster processing per task, and more uniform cycle counts across all kernels contribute to improved overall system throughput from a task pipeline perspective.

Compared to the Naive method, the Stream and Window methods significantly reduce cycle counts for each kernel, demonstrating the effectiveness of In-tile optimizations, particularly Vector Optimization. Additionally, task partitioning further improves pipeline efficiency by splitting the color and cov2D kernels into the dir\_vec and Jacobian kernels. 
However, additional performance gains may require further partitioning of the color kernel to better distribute the workload.

\subsubsection{Evaluation of the Entire System}

\begin{table}[htbp]
  \centering
  \caption{Comparison of performance metrics for different methods, including Feature Computation Time and Throughput.}
  \label{tab:performance_comparison}
  \scriptsize
  \renewcommand{\arraystretch}{1.2} 
  \setlength{\tabcolsep}{3pt} 
  \resizebox{\linewidth}{!}{ 
  \begin{tabular}{|c|c|c|c|c|c|}
  \hline
  \textbf{Metric} & \textbf{Non-AIE} & \textbf{Naive} & \textbf{Stream} & \textbf{Window-1} & \textbf{Window-25} \\ \hline
  Feature Computation Time (s) & 0.278 & 0.780 & 0.393 & 0.399 & 0.389 \\ \hline
  Throughput (MB/s) & 64.1 & 22.9 & 45.3 & 44.7 & 45.8 \\ \hline
  \end{tabular}
  }
\end{table}

Table~\ref{tab:performance_comparison} presents the results of measuring the feature computation time and throughput for Gaussian computations on the VCK190 hardware.
Additionally, the total execution time, defined as the time from system initialization to the completion of image generation, is also reported.

The Non-AIE case represents the scenario where Gaussian feature computation was performed solely on the PS of the VCK190 without utilizing the AI Engines.
Naive, Stream, and Window represent cases where the AI Engines were employed.
Naive is a baseline implementation without any optimizations, whereas Stream and Window incorporate in-tile optimizations within the AI Engines.
Window-1 represents the use of the Window interface with a single feature computation unit, while Window-25 represents the use of 25 feature computation units in parallel.

The experimental results, contrary to expectations, showed that the Non-AIE case exhibited the highest efficiency.
Stream and Window demonstrated approximately twice the throughput compared to Naive, yet they still fell short of the performance achieved by the Non-AIE case.
Furthermore, the throughput of Stream, Window-1, and Window-25 all saturated at approximately 45 MB/s, and unlike the AI Engines simulation results, increasing the degree of parallelism had little impact on improving throughput.

PLIO provides an overall bandwidth of 1.0 TB/s, and the DDR Memory Controller (DDRMC) offers a bandwidth of 25.6 GB/s.
Based on these values, it is considered that sufficient bandwidth is secured for data transfer within the AI Engines.
Therefore, it is highly likely that the performance bottleneck lies not in the AI Engine's computational performance, but in the data transfer process within the PL.
Since the AI Engines simulation did not account for data transfer processes within the PL, discrepancies arose between the actual hardware results and the simulation results.

\section{Conclusion}

In this study, we proposed a hardware algorithm to accelerate feature computation for 3DGS on the Versal AI Engine by fully leveraging the capabilities of the AI Engines. 
A comprehensive evaluation of the proposed approach revealed several important findings.

First, significant throughput improvements were achieved through AI Engines optimization. 
Compared to the Naive implementation, the use of In-tile optimizations with Stream and Window interfaces substantially increased the throughput of Gaussian feature computation. 
Furthermore, by exploiting the spatial parallelism of AI Engine Array, the throughput was enhanced by up to 226 times over the Naive implementation. 
These optimization techniques are not only effective for 3DGS but are also expected to be applicable to other high-performance computing applications.

Second, the study identified bottlenecks in system-wide execution. 
While the hardware algorithm leveraging the AI Engines demonstrated theoretical throughput improvements in simulation environments, practical results on the actual hardware revealed that data transfer processing within the PL significantly limited performance. 
This bottleneck hindered the full realization of the expected throughput improvements, emphasizing the need for more efficient data transfer mechanisms in the PL to fully exploit the capabilities of the AI Engines.

\section*{Acknowledgment}
This work is supported in part by JST CREST JPMJCR21D2 and the research collaboration with Konica Minolta.

\bibliographystyle{ieeetr}
\bibliography{bibliography}

\begin{thebibliography}{1}

\bibitem{theis2017end}
T.~N. Theis and H.-S.~P. Wong, ``The end of moore's law: A new beginning for
  information technology,'' {\em Computing in science \& engineering}, vol.~19,
  no.~2, pp.~41--50, 2017.

\bibitem{gaide2019xilinx}
B.~Gaide, D.~Gaitonde, C.~Ravishankar, and T.~Bauer, ``Xilinx adaptive compute
  acceleration platform: Versaltm architecture,'' in {\em Proceedings of the
  2019 ACM/SIGDA International Symposium on Field-Programmable Gate Arrays},
  pp.~84--93, 2019.

\bibitem{kerbl20233d}
B.~Kerbl, G.~Kopanas, T.~Leimk{\"u}hler, and G.~Drettakis, ``3d gaussian
  splatting for real-time radiance field rendering.,'' {\em ACM Trans. Graph.},
  vol.~42, no.~4, pp.~139--1, 2023.

\bibitem{xilinx_aie_api}
``Ai engine api user guide: Api reference.''
  \url{https://www.xilinx.com/htmldocs/xilinx2022_2/aiengine_api/aie_api/doc/modules.html}.

\bibitem{xilinx_aie_intrinsics}
``Ai engine intrinsics user guide: Overview.''
  \url{https://www.xilinx.com/htmldocs/xilinx2023_2/aiengine_intrinsics/intrinsics/index.html}.

\bibitem{amd_datasheet}
``Versal^^e2^^84^^a2 architecture and product data sheet: Overview (ds950).''
  \url{https://docs.amd.com/v/u/en-US/ds950-versal-overview}.

\bibitem{mildenhall2021nerf}
B.~Mildenhall, P.~P. Srinivasan, M.~Tancik, J.~T. Barron, R.~Ramamoorthi, and
  R.~Ng, ``Nerf: Representing scenes as neural radiance fields for view
  synthesis,'' {\em Communications of the ACM}, vol.~65, no.~1, pp.~99--106,
  2021.

\bibitem{schonberger2016structure}
J.~L. Schonberger and J.-M. Frahm, ``Structure-from-motion revisited,'' in {\em
  Proceedings of the IEEE conference on computer vision and pattern
  recognition}, pp.~4104--4113, 2016.

\bibitem{inria_gaussian_splatting}
I.~G. Team, ``3d gaussian splatting dataset.''
  \url{https://github.com/graphdeco-inria/gaussian-splatting}.

\end{thebibliography}
\end{document}